\documentclass[pr,twocolumn,showpacs]{revtex4}
\usepackage{bm}
\usepackage{graphicx}
\usepackage{amsmath}
\usepackage{eufrak}
\usepackage{color}
\usepackage{upgreek}
\usepackage{subfigure}
\usepackage[unicode=true,colorlinks=true,citecolor=blue]{hyperref}
\usepackage{placeins}
\usepackage{lipsum}

\usepackage[english]{babel}

\begin{document}

\title{Terahertz electric field driven electric currents and ratchet effects in graphene}

\author{Sergey D.~Ganichev, Dieter Weiss, and Jonathan Eroms,  }
\affiliation{Terahertz Center, University of Regensburg, 93040 Regensburg, Germany}
\begin{abstract}
Terahertz field induced photocurrents in graphene were studied experimentally and by microscopic modeling. Currents were generated by $cw$ and pulsed laser radiation in large area as well as small-size exfoliated graphene samples. We review general symmetry considerations leading to photocurrents depending on linear and circular polarized radiation and then present a number of situations where photocurrents were detected. Starting with the photon drag effect under oblique incidence, we proceed to the photogalvanic effect enhancement in the reststrahlen band of SiC and edge-generated currents in graphene. Ratchet effects were considered for in-plane magnetic fields and a structure inversion asymmetry as well as ratchets by non-symmetric patterned top gates. Lastly, we demonstrate that graphene can be used as a fast, broadband detector of terahertz radiation. 
\end{abstract}
\maketitle

\section{Introduction}

 {The advent of graphene and topological insulators (TI) started a new research direction in materials science.} A distinctive feature of these materials is that their band structure resembles the dispersion relation of a massless relativistic particle being described by the Dirac equation. 
 {Transport effects linear in electric field have been studied extensively in those materials, leading to significant progress both in 
basic research and a number of applications (see, e.g.~\cite{Neto,Peres,Avouris,Sarma,Young,McCann} for a review).}
Unique optical properties of this material also caused a  rapid development of graphene photonics and optoelectronics, see e.g.~\cite{Bonaccorso,mueller,novoselov}.
{Nonlinear transport effects, being proportional to higher powers of the field,} offer a new playground for many interesting phenomena in the physics of Dirac fermions (DF)~\cite{GlazovGanichev}. 
{These effects are usually} caused by the radiation induced redistribution of charge carriers in momentum/energy space and reconstruction of the energy spectra. 
The resulting response  {comprises components which oscillate in time and space, but also has a steady-state and spatially uniform contributions.} 
{Therefore, both $ac$ and $dc$ currents are generated, consisting of terms whose magnitudes depend non-linearly on the field amplitude and which are controlled by the radiation polarization.}

Edge and bulk photocurrents have been detected in many DF systems excited by infrared/terahertz radiation, giving insights into the microscopic mechanisms and the requirements for such kind of experiments. Focusing on DFs in graphene, some recent theoretical and experimental examples of such phenomena include the circular dynamic Hall effect \cite{karch2010,2010arXiv1002.1047K}, circular and linear photogalvanic effects\cite{2010arXiv1002.1047K,jiangPRB2010,reststr2013}, chiral edge photocurrents~\cite{edgePRL2011}, coherent current injection~\cite{sun2010,Sun:2012ys,2012winzent}, magnetic quantum ratchet\cite{Drexler2013}, ratchet effects with lateral potential~\cite{ratchet_graphene16}, time-resolved photocurrents~\cite{Prechtel:2012kx,Graham:2013uq,detectorFEL1,detectorFEL}. For a review on non-linear electron transport in bulk graphene at $B=0$ see~\cite{GlazovGanichev}. 
These  {studies demonstrate substantial differences of the microscopic mechanisms of non-linear transport effects in DF systems and conventional semiconductors.}
Thus, the experimental and theoretical research in the field of non-linear optics and optoelectronics in DF in infrared/terahertz spectral range becomes already an important task, for reviews 
see~\cite{GlazovGanichev,plasmawave,Freitag,koppens,Hartmann,Tredicucci14}. Furthermore, infrared/THz spectroscopy turns out to be an efficient tool providing information on band parameters, Fermi velocity, symmetry properties, carrier dynamics, etc. From an application point of view, converting an $ac$ electric THz field into a $dc$ current is a very promising route towards fast, sensitive detection of terahertz radiation at room temperature~\cite{detectorFEL1,detectorFEL}. 

In this feature article, we will focus on the effects of the lowering of symmetry by various mechanisms on photo-induced currents in graphene. Due to the high symmetry of graphene, such signals are forbidden for normal incidence of the electromagnetic radiation. By changing the angle of incidence or lowering the symmetry, signals are obtained. 

We will give an overview of the terahertz radiation induced photocurrents in graphene, 
{describe principal experimental and theoretical findings of non-linear physics in graphene and suggest further studies in this research area.}
{We will first briefly introduce the methods} used to study nonlinear phenomena in graphene. Then, in sections \ref{drag}-\ref{edge} we describe 
photocurrents generated in pristine graphene and at graphene edges.
In Secs. \ref{magnet}-\ref{ratchet} we address ratchet effects in graphene. We begin with ratchet effects caused by the application of an external magnetic  field caused by the periodic  radiation field and structure inversion asymmetry. Then we describe 
ratchet effects in graphene superimposed with periodic asymmetric lateral potential.
{For each effect, we will proceed in the following way: We present symmetry arguments allowing a phenomenological analysis of the respective phenomena, then outline the microscopic theory and finally discuss the main experimental findings.}
In Sec.~\ref{detector} we discuss the application of graphene
photoelectrical phenomena   for fast room temperature detection of infrared/terahertz radiation. Finally, in Sec.~\ref{summary} we summarize the results and discuss the prospects of future theoretical and experimental studies of the nonlinear electromagnetic response of graphene. 

\section{Methods}
\label{methods}

\begin{figure*}
  \includegraphics[width=0.9\textwidth]{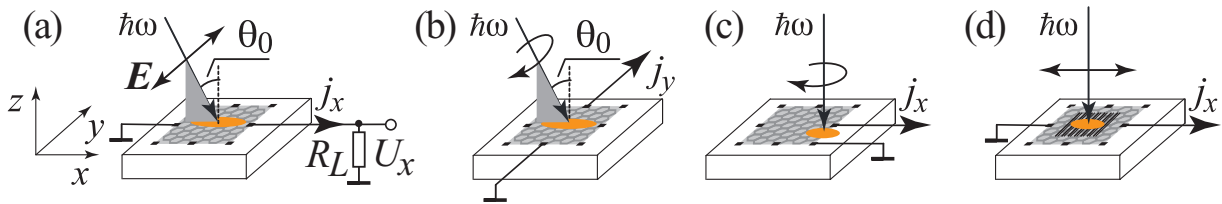}%
  \caption{\label{figure1experimental} 
 {Measurement configurations for the detection of longitudinal (a) and transverse (b) photocurrents. The plane of incidence of the radiation is also defined. Black dots: Contacts to graphene.}
(c) and (d) demonstrate illumination with circularly and linearly polarized light.}
\end{figure*}

\subsection{Symmetry analysis}
\label{symmetry}

Photocurrents discussed in this paper are phenomenologically
described by writing the current  as an expansion in powers of
the electric field  $\bm E = \bm E(\omega) \exp{(-i\omega t)} + {\rm c.c.}$
at the frequency $\omega $ and the wavevector $\bm q$ 
of the radiation field inside the 
medium~\cite{GlazovGanichev,sturmanBOOK,ivchenkopikus,ivchenko05a,ganichev_book}. 
The lowest order nonvanishing terms yielding a $dc$ current density
$\bm j$ are given by
\begin{equation}
\label{pgepdegeneral1} j_{\lambda} =\sum_{\mu, \nu}
\chi_{\lambda \mu \nu} E_{\mu} E^*_{\nu} + \sum_{\delta, \mu, \nu}
T_{\lambda \delta \mu \nu} q_{\delta} E_{\mu} E^*_{\nu} \:,
\end{equation}
where $E^*_{\nu} = E^*_{\nu}(\omega ) = E_{\nu}(-\omega )$ is  the
complex conjugate of $E_{\nu}$. 
The expansion coefficients $\chi_{\lambda \mu \nu}$ and
$T_{\lambda\mu \nu \delta}$ are third rank and fourth rank
tensors, respectively. The first term on the right-hand side of
Eq.~(\ref{pgepdegeneral1}) represents photogalvanic (PGE)
effects.  {The second term} containing the wavevector of the electromagnetic field describes
the photon drag (PDE) effect.
Both effects are sensitive to the radiation polarization 
which is defined by the variation of product $E_{\mu} E^*_{\nu}$.
In general both, photogalvanic and photon drag effect, 
yield photocurrents depending on the degree of linear polarization
and on the radiation helicity as well as have a contribution
being independent of the radiation polarization.
Taking the example of the photogalvanic effects, we 
obtain the photocurrent contributions attributed to the action of
linearly and circularly polarized radiation.
The bilinear combination $E_\mu E^*_\nu$ can be rewritten as a sum of a
symmetric and an antisymmetric  product
\begin{equation}
\label{pgepdegeneral2} E_\mu E^*_\nu = \{ E_\mu E^*_\nu\} +
[E_\mu E^*_\nu] ,
\end{equation}
with
\begin{equation}
\label{pgepdegeneral3} \{ E_\mu E^*_\nu\} = \frac{1}{2}
(E_\mu E^*_\nu + E_\nu E^*_\mu ) 
\end{equation}
and
\begin{equation}
\label{pgepdegeneral3b} 
[E_\mu E^*_\nu ] = \frac{1}{2} (E_\mu E^*_\nu - E_\nu E^*_\mu ).
\end{equation}

This decomposition  of $E_\mu E^*_\nu$ corresponds to a splitting
into  real and  imaginary parts. The symmetric term is real while
the antisymmetric term  is purely imaginary. Due to contraction of
the tensor $\chi_{\lambda\mu \nu}$ with $E_\mu E^*_\nu$ the same
algebraic symmetries  are projected onto the last two indices of
$\chi_{\lambda\mu \nu}$. The real part of $\chi_{\lambda\mu \nu}$
is symmetric in indices $\mu\nu$ whereas the imaginary part is
antisymmetric. Antisymmetric tensor index pairs can be reduced to
a single  pseudovector index using the Levi--Civita totally
antisymmetric tensor $\delta_{\rho\mu\nu}$. Applying this
simplification  we obtain for the current due to the antisymmetric
part of $E_\mu E^*_\nu$
\begin{equation}
\label{pgepdegeneral4} \chi_{\lambda\mu\nu}[E_\mu E^*_\nu ] =
i\cdot \sum_{\rho}\gamma_{\lambda\rho}\delta_{\rho\mu\nu}[E_\mu
E^*_\nu ] = \gamma_{\lambda\rho}i(\bm E \times \bm E^*)_\rho ,
\end{equation}
with  the real second rank pseudotensor $\gamma_{\lambda\rho}$ and
$i(\bm E\times \bm E^*)_\rho =
\hat{e}_\rho P_{\rm circ}\:E^2$, where $\bm \hat{e} = \bm q /q $, $P_{\rm circ}$ and $E^2 = |E(\omega)|^2$  are the unit
vector pointing in the direction of light propagation, 
degree of light circular polarization (helicity) and the radiation intensity,
respectively. In summary we find for the total photogalvanic current
\begin{equation}
\label{pgepdegeneral5} j_{\lambda}^{\rm PGE} =\sum_{\mu, \nu}
\chi_{\lambda \mu \nu} \{E_{\mu} E^*_{\nu}\} + \sum_{\rho}
\gamma_{\lambda \rho}\:i (\bm E \times \bm E^*)_{\rho} 
\:,
\end{equation}
where $\chi_{\lambda\mu \nu} = \chi_{\lambda \nu\mu}$. In this
equation  the photogalvanic effect is decomposed into the LPGE
(linear photogalvanic effect)
and the CPGE (circular photogalvanic
effect)
described by the first and second term on the right-hand side, respectively. 
{The corresponding contributions for the photon drag current 
can be obtained in a similar way. }

Linear and circular photogalvanic and photon drag 
currents have been observed in various semiconductors and are
theoretically well understood (for reviews see,
e.g.~\cite{GlazovGanichev,sturmanBOOK,ivchenkopikus,ivchenko05a,ganichev_book,Ch7Yaroshetskii80p173,Ch7Gibson80p182,IvchenkoGanichev,GanichevGolub}).

 {Symmetry analysis permits us to describe the various effects and their observability in terms of macroscopic parameters, such as radiation intensity, polarization and angle of incidence without detailed knowledge of the microsopic origin. Disregarding the substrate, a homogeneous,  infinite}  {pristine}  {graphene layer belongs to the centrosymmetric $D_{6\rm h}$ point group. When a substrate or asymmetrically placed adatoms are present, the symmetry is reduced to the the noncentrosymmetric  group $C_{6\rm v}$, removing the equivalence of the  $z$ and $-z$ directions.}
While the photon drag effect can be detected for both kinds of graphene structures
photogalvanic effects, which require the lack of inversion symmetry, 
can not be excited in an infinite homogeneous pristine graphene layer. 
Symmetry analysis shows that photocurrents in the graphene systems addressed above 
can be generated for oblique incidence only and may have a
contribution along radiation propagation and 
normal to it, see Fig.~\ref{figure1experimental}
(a) and (b). 
However, in real structures photocurrents excited by perpendicularly incident radiation 
may become become possible,  {e.g. when the edges are illuminated}, see Fig.~\ref{figure1experimental}(c), 
 {and symmetry is reduced locally,} 
or in samples with ripples or terraces.  
The photocurrent at normal incidence may further become possible for graphene with 
asymmetric metal structures on its top, see Fig.~\ref{figure1experimental}
(d), or  {when an plane static magnetic field is applied.}
 {Since second-order phenomena are sensitive to spatial inversion, particular properties of the samples, like the presence of adatoms, terraces, ripples and edges, or the coupling to the substrate become important.}

 {Furthermore, those effects depend strongly on  angle of incidence and the 
radiation polarization.}
Studying of these dependencies, together with the symmetry analysis, 
helps to explore microscopic mechanisms responsible for the 
photocurrent generation.  Our works demonstrate
that the various possible contributions to the nonlinear response are just 
proportional to the Stokes parameters, which describe the polarization state of radiation.
 {Hence, when performing measurements of nonlinear high frequency effects in graphene, we}  {vary} 
 {the radiation polarization state}
by rotating standard 
dichroic elements like $\lambda/2$ and $\lambda/4$ 
plates or Fresnel rhombs with respect to the polarization plane 
of the linearly polarized laser radiation with ($\bm E_l\parallel x$). 
 {For}
light propagating  {in the direction of} 
the positive $z$ axis, the Stokes parameters~\cite{Saleh,BornWolf} are given by
\begin{equation}
\label{S2}
S_1 = \frac{|E_x|^2-|E_y|^2}{|E_x|^2+|E_y|^2}\propto \cos{2\alpha}\propto \cos^2{2\varphi}, 
\end{equation}
\begin{equation} 
\label{S2}
S_2 = \frac{E_xE_y^*+E_x^*E_y}{|E_x|^2+|E_y|^2} \propto \sin{2\alpha} \propto \sin{4\varphi},
\end{equation}
\begin{equation}
\label{S3}
S_3 \equiv P_{\rm circ}= \mathrm i \frac{E_xE_y^*-E_x^*E_y}{|E_x|^2+|E_y|^2} = \sin{2\varphi} \, ,
\end{equation}
where $|E_x|^2+|E_y|^2$ defines the radiation intensity, 
$\alpha=2\beta$ is the azimuth angle defining the orientation of the  polarization plane for linearly polarized radiation
and $\beta$ and $\varphi$ are the angles between $\bm E_l$  the optical axis $c$ for 
half- and quarter wave elements, respectively. 

\subsection{Experimental}
\label{experimental}

Photocurrents in graphene have been observed and studied applying 
radiation from near- up to very far-infrared range. 
To cover a wide frequency range, stretching over three 
decades from fractions- up to tens of terahertz, various radiation sources have been applied including molecular optically pumped $cw$ and pulsed lasers at Regensburg Terahertz Center (for laser characteristics see e.g.~\cite{DMS2009,Kvon2008,Tunnel1993,DX1995,spinrelax,Tunnelingfrequency1998}), 
free electron lasers Felbe in Rossendorf~\cite{detectorFEL1,detectorFEL,FEL1,FEL2} and Felix in the Netherlands~\cite{Knippels99p1578,weber2008}, tunable CO$_2$ lasers~\cite{ganichev_book}, quantum cascade lasers~\cite{Faist1994,Faist2000,helm2001} and backward wave oscillators~\cite{Bruenderman2012}. Using of various sources of radiation not only allowed to 
explore frequency dependencies of the photocurrents under study, but 
also  made  possible demonstration of they robustness to high radiation power
and examination of the subnanosecond photocurrent dynamics.
We emphasize that the photocurrents are detectable at very low power of microWatts and yield 
response linearly scaling with radiation power up to at least 100 kW without samples
damage.

The main experimental geometries used for these studies are  {outlined} in Fig.~\ref{figure1experimental}. 
 {We illuminated the graphene samples under normal or oblique incidence,}
with the incidence angle $\theta_0$  varying from $-40^\circ$ to +40$^\circ$. 
The photocurrents have been measured as a voltage drop across a load resistance, $R_L$.
For experiments with pulsed lasers the photovoltage signal, $V$, was detected by a digital oscilloscope 
and for $cw$ radiation modulated with a chopper  by using  standard lock-in technique.
While for the measurements applying pulsed lasers with nanoseconds pulse duration
$R_L = 50$~Ohm has been used for $cw$ radiation 
in some cases much higher load resistance have been used {as well}. 
In the former set-up the photocurrent $I$ relates to the  photovoltage $V$ as $I  =  V / R_L$,  
because in all experiments described below the load resistance was 
much smaller than the sample resistance $R_S$ ($R_L \ll R_S$). The corresponding photocurrent density is obtained as 
$j = I / w$, where $w$ is the width  of the graphene sample. 
The latter configuration greatly increases the magnitude of the voltage signal but 
complicates the analysis for the case when sample resistance varies 
substantially with an external parameter, e.g magnetic field or gate voltage.
 {The radiation was focused 
onto the samples by a parabolic mirror and its power $P$ was}  {controlled by  photon drag and 
pyroelectric detectors.}
The  beam shape of the THz radiation is almost Gaussian, 
 {measured}  
with a pyroelectric camera~\cite{Ganichev1999,Ziemann2000}. 

Photocurrents of different microscopic origins have been observed in a
large temperature range from 2 up to 300~K. We emphasize that all
observed photocurrents have been {also} detected at technologically 
important room temperature. While studying of the temperature dependence
also played an important role for understanding the photocurrent 
formation the largest portion of the research was focused on 
room temperature response.  

\subsection{Samples}
\label{samples}

Terahertz radiation induced photocurrents have been observed and studied
in graphene samples prepared applying different technologies including: 
(i) epitaxial graphene prepared by high temperature Si sublimation of semi-insulating silicon carbide (SiC) substrates~\cite{Bostwick09,Virojanadara08,Emtsev2009,Seyller2,Tzalenchuk2010,suppllara2011,SwedenPRL2011}, (ii) CVD graphene grown in a conventional chemical vapor deposition process using
copper as substrate and catalyst and methane as carbon source~\cite{Drexler2013} and (iii) 
exfoliated graphene~\cite{Bib:Novoselov2004} deposited on oxidized silicon
wafers. Details on the  growth  and characterization of material used for photocurrent studies
can be found in~\cite{karch2010,edgePRL2011,Drexler2013,ratchet_graphene16,detectorFEL1,detectorFEL,Emtsev2009,Tzalenchuk2010,suppllara2011,SwedenPRL2011}. 
The technologies (i) and (ii) allowed us to prepare large area 
samples with $5 \times 5$~mm$^2$ graphene monolayers while the size of the exfoliated
structures was in the range of tens of micrometers. 
The large size of the epitaxial and CVD samples was of particular importance for the
analysis of the photocurrent formation.  While both large and small size 
samples showed the effects, the response of the micron
sized exfoliated samples in all type of experiments had an unavoidable 
contribution of the edge photocurrents discussed in Sec.~\ref{edge}.
This is because the spot size of the terahertz laser of about
1~mm$^2$ is much larger than the graphene flakes. 

Most experiments were carried out on \textit{n}- and \textit{p}-type layers 
with carrier concentrations in the range of ($0.5$ to $7$)$\times$10$^{12}$ cm$^{-2}$, 
Fermi energy $E_F$ of several hundreds of meV, 
and mobilities about 1000\,cm$^2$/Vs at room temperature.
We note that in all photocurrent experiments described below 
 {were performed in the limit $\hbar\omega \ll E_F$.}  {Thus microscopic theory of the studied 
phenomena has been developed for the \emph{classical} regime of light-matter interaction.}
Thus a  microscopic description of each photocurrent under study was obtained by solving the Boltzmann kinetic equation for the electron distribution function $f(\bm p, \bm r,t)$.  {Here $\bm p$ is 
 the free-carrier momentum, $\bm r$ is the in-plane coordinate  and  $t$ is time.}

Electron transport parameters have been obtained from magneto-transport measurements.
For some experiments, e.g. on terahertz ratchet effects described in Sec.~\ref{ratchet}, 
concentration and type of carriers have been controllably changed 
applying top and back gate voltages. 

To obtain defined graphene edges  an edge trim  of about 200~$\mu$m width
was  removed by reactive ion etching with an argon/oxygen 
plasma.
To protect graphene from uncontrollable change of transport parameters most of 
samples were encapsulated in a polymer film~\cite{suppllara2011}, consisting of 
 {PMMA/MAA thin film followed by ZEP520 polymer.}
The unprotected samples were  {subject} to contamination from the ambient
atmosphere. The latter has been seen from the  {change in}  carrier  {mobility and density} on a
time scale of months.

For electrical measurements eight  {contacts} were  {made
in the corners and at the middle}
of the sides of the square shaped large size graphene layer.  {The contacts have been fabricated
by} e-beam deposition of 3~nm Ti and 80~nm Au  {using} a laser-cut shadow mask. Each of the electrodes had
200x200~$\mu$m$^2$ lateral dimension. Raman spectra taken from several points of each
of the samples showed high crystallinity, 1-2 atomic layer thick graphene. Metal contacts to 
graphene flake has been prepared on the periphery of graphene  {applying}
 {standard lithographic} deposition of Ti/Au (3/100~nm) and lift-off.

For studying ratchet effects, metal film superlattices were fabricated on large area epitaxial graphene as well as on small area flakes. 
A sketch of the superlattices gate fingers and a corresponding optical micrograph are 
presented in Sec.~\ref{ratchet}. {Preparing gates,} first, an insulating aluminium oxide layer was deposited  on  top of the graphene sheet. 
The lateral periodic electrostatic potential is created on top of epitaxial graphene by periodic grating-gate fingers  fabricated by
electron beam lithography and subsequent deposition of metal (5~nm Ti and 60 nm Au).
On small area graphene flakes
we fabricated inter-digitated metal-grating gates (5 nm/60 nm  Ti/Au) TG1 and TG2 
having  different stripe width and stripe separation. This allowed us to apply different 
bias voltages to the individual subgrating gates forming the superlattice.
The samples were glued onto holders with conductive epoxy utilizing
the highly doped silicon wafer as a back gate
which enabled us to  change type and density of free carriers in graphene.
Contact pads were placed in a way that the photo-induced currents 
can be measured parallel  and perpendicular 
to the metal fingers.

\section{Photon drag effect}
\label{drag}

\begin{figure}
  \includegraphics[width=0.9\columnwidth]{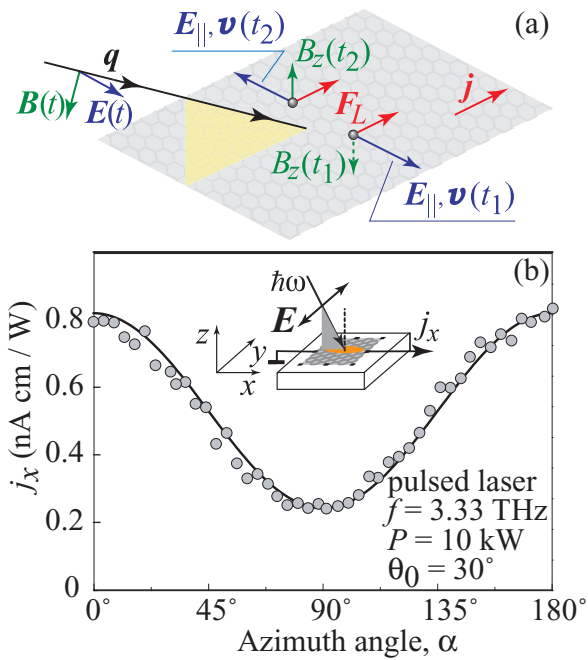}%
  \caption{\label{figure2hall1}
   (a)     {Sketch illustrating the dynamic Hall effect assuming positively charged carriers for clarity}. 
 {Here	$\bm E_\parallel$ and  $B_z$ are
the radiation in-plane component of electric field and $z$-component of the magnetic field, respectively.
$\bm v$ is the electron velocity induced by electric field of the radiation.} 
These vectors are shown for two  {moments in time}, $t_1$ and $t_2$,  {separated by half a period} of the field oscillations. Microscopically, action of these fields results in $\bm F_L$ and, correspondingly, $\bm j$ are the Lorentz force and $dc$ current, respectively, see text for details. 
(b) Longitudinal photon drag current as a function of the azimuth angle $\alpha$
defining orientation of the radiation electric field vector.
Data are given after~
\cite{karch2010,GlazovGanichev}.
}
\end{figure}

We demonstrated the photon drag effect in graphene produced both by exfoliation and epitaxial techniques~\cite{karch2010,2010arXiv1002.1047K,jiangPRB2010,GlazovGanichev} under oblique incidence of radiation. Electric currents were obtained both in the direction of the radiation propagation (longitudinal geometry) as well as perpendicular to it (transverse geometry). 
Using the Boltzmann kinetic equation 
we obtain a microscopic description of the photon drag:
\begin{equation}
 \label{kinetichall}
\frac{\partial f}{\partial t} + \bm v \frac{\partial f}{\partial \bm r} + e \left(\bm E + \bm v \times \bm B\right)
\frac{\partial f}{\partial \bm p} = Q\{f\}\:.
\end{equation}
Here, $\bm{v} = d \varepsilon / d \bm{p}$ is the velocity,  $\varepsilon$
is the kinetic energy, and $Q\{f\}$ is the collision integral described in terms of
relaxation times $\tau_n$ ($n=1,2\ldots$) for
corresponding angular harmonics of the distribution function~\cite{2010arXiv1002.1047K,GlazovGanichev,perelpinskii73}. The electric current density is given by the standard equation 
\begin{equation}
 \label{generalcurrent}
\bm j = 4 e \sum_{\bm p} {\bm v}\, f(\bm p) \,, 
\end{equation}
where $e$ is the electron charge and a factor of 4 accounts for  spin and valley degeneracies. 
We expand the distribution function in powers of
electric and magnetic fields,  {retaining} linear and quadratic terms only. 
Calculations of $f(\bm p)$ and ${\bm j}$ are carried out using the
energy dispersion $\varepsilon_p= \pm v p$ of free carriers in
graphene and the relation $\bm v \equiv \bm v_{\bm p}=v \bm p/|\bm
p|$ between the velocity and the quasi-momentum ($v \approx c/300$,
with $c$ being the speed of light). 
The photon drag current can be generated either due to combined 
action of the electric and magnetic field of the electromagnetic wave ($EB$-mechanism or the dynamic Hall effect) or due to the spatial gradient of the in-plane projection of the radiation electric field ($qE^2$-mechanism). 
 {Since in plane waves  the complex
amplitudes of electric and magnetic fields}  {in Eq.~(\ref{kinetichall})}  {are coupled, $\bm B(\omega,\bm q) =\frac{1}{c|\bm q|} [\bm q \times \bm E(\omega,\bm q)]$ (taking 
$\epsilon_0=1$ and $\mu_0=1$), both mechanisms share the same origin. 
Therefore, the dynamic Hall effect  $\propto E_\beta B_\gamma^*$ can be expressed in terms of the photon drag effect, i.e. $\propto q_\delta E_\beta E_\gamma^*$. When the effect is treated microscopically in terms of the number of photons (quantum mechanical picture), we speak of the photon drag effect, while the classical picture using the action of electromagnetic fields results in the term dynamic Hall effect.}

The  {model of} 
the dynamic Hall effect excited by linearly polarized radiation is illustrated in Fig.~\ref{figure2hall1}(a). 
At one given moment in time, $t_1$, the 
 {Lorentz force caused by the radiation electric and magnetic fields results in}
a drift in the direction of the light propagation. Half a radiation period later, at time $t_2$, both fields have changed their sign, therefore, the drift direction remains. Averaging over time, this leads to a time-independent Hall current with fixed direction.  {The latter depends on the electric field vector orientation and is odd in the 
angle of incidence, $\theta_0$}. 
The additional contribution caused by the $qE^2$-mechanism is {also} odd in $\theta_0$ and vanishes for normal incidence. 
 {The terms of the fourth-rank
tensor $T_{\alpha\beta\gamma\mu}$ which are symmetric and antisymmetric under $\beta\gamma \leftrightarrow \gamma\beta$, yield a photon drag contribution responding to linearly and circularly polarized radiation, respectively (in short: linear and circular photon drag effects)~\cite{Ivchenko1980,belinicher_cpde,Shalygin2007,PhysRevLett.103.103906}.}
While the longitudinal current can be understood intuitively for arbitrary polarization, the transverse current obtained by circularly polarized radiation is not obvious. It changes its sign upon reversing the helicity of the radiation, and retardation the electric field $\bm E$ and the instant velocity of charge carrier $\bm v$ has to be taken into account~\cite{karch2010}. It is most pronounced for $\omega\tau \sim 1$.  
Now, in the  {schematic model} of Fig.~\ref{figure2hall1},  the carriers 
{will} follow an elliptic orbit instead of a linear trajectory. Due to retardation,  the velocity ${\bm v}$ does not immediately track the  {instantaneous} ${\bm E}_{\vert \vert }$-field direction.  {Instead, a} phase shift equal to $\arctan(\omega \tau)$ between the electric field and the electron velocity ${\bm v}$  {ensues.}
Ultimately, this results in a $y$-component of  the Lorentz force $\bm F_L$, which depends on the direction of electron motion and, consequently, on the radiation helicity. 
The microscopic theory for the photon drag effect in graphene was developed 
in Refs.~\cite{karch2010,2010arXiv1002.1047K} for  {the} classical frequency range and 
in Ref.~\cite{jiangPRB2010,GlazovGanichev} for the quantum frequency range.

\begin{figure}
  \includegraphics[width=\columnwidth]{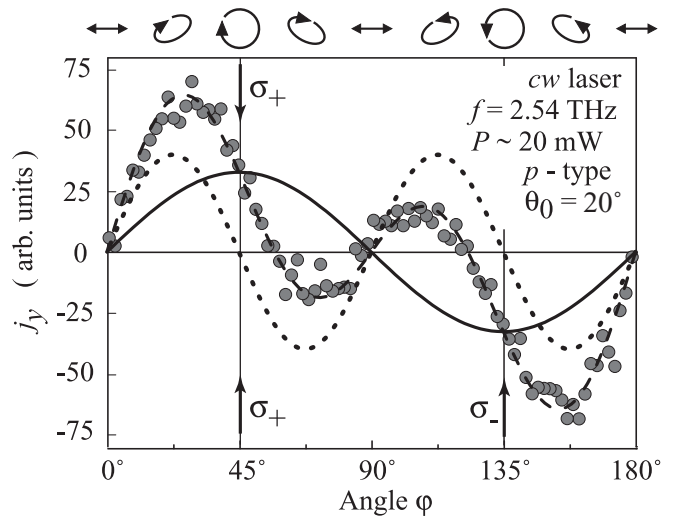}%
  \caption{\label{figure3hall2}
 {Dependence of the photocurrent $j_{y}$ on }
the angle $\varphi$.
 {The polarization states for various $\varphi$ are illustrated by ellipses (top).} Dashed lines: fits to $j_{y} = J_A+j_B = A \theta_0 \sin 2\varphi + B \theta_0 \sin 4\varphi$  {including} the circular contribution $j_A$ (full  line) and the linear contribution $j_B$ (dotted line). Data are given after
\cite{karch2010}.
}
\end{figure}

Experimentally, the photon drag effect, including dynamic Hall effect contribution, was demonstrated both in exfoliated and epitaxial graphene samples for a wide frequency range from fractions of terahertz up to tens of THz~\cite{karch2010,2010arXiv1002.1047K,jiangPRB2010,GlazovGanichev}. 
 {We used highly resistive Si or semi-insulating SiC substrates to rule out high losses or shunting by conductive substrates.}
Due to the mm-size diameter of the Gaussian beam, illumination of the sample edges could not be avoided in the exfoliated samples. This leads to an additional edge current contribution, which is covered in Sec~\ref{edge}. 
Circular and linear photon drag effects have been observed 
in a wide range of temperatures (from room temperature down to liquid helium temperature)
in both \textit{n}- and \textit{p}-type layers with carrier concentrations
in the range of ($3$ to $7$)$\times$10$^{12}$ cm$^{-2}$ and mobilities about 
1000\,cm$^2$/Vs at room temperature, and in a 
 wide range of radiation 
intensities, from mW/cm$^2$ up to MW/cm$^2$.

An example of the polarization dependence of the longitudinal photon drag effect is shown in Fig.~\ref{figure2hall1}(b). Figure~\ref{figure3hall2} shows  {transverse photocurrent excited by elliptically polarized radiation in epitaxial single layer graphene.} 
The rotation angle $\varphi$ of the quarter-wave plate controls the polarization state of light.  {The figure reveals that}
the photocurrent
 {is composed of  circular  and linear terms of comparable strength. }
Changing from left- to right-handed circular polarization, the circular  contribution ($\bm j \propto P_{\rm circ} = \sin{2\varphi}$) changes its sign. In transverse direction, we observe the both the linear and circular contribution, while the signal detected in the incidence plane consists of a linear contribution together with polarization independent current, in agreement with symmetry arguments. The microscopic theory yields
\begin{subequations}
\label{j:phenHall}
\begin{equation}
\label{j:xHall}
j_x/E^2 = T_1  q_x \frac{|e_x|^2 + |e_y|^2}{2}
+ T_2 q_x \frac{|e_x|^2 - |e_y|^2}{2}  ,
\end{equation}
\begin{equation}
\label{j:yHall}
j_y / E^2 = T_2 q_x \frac{e_xe_y^* + e_x^*e_y}{2}  - \tilde{T}_1  q_x P_{\rm circ} \hat{e}_z .
\end{equation}
\end{subequations}
where {$T_1, T_2, T_3$ and $\tilde{T}_1$ denote linearly independent components of the tensor $T_{\lambda \delta \mu \nu}$,} $x$ and $y$ are the axes in the graphene plane, and $z$ is
the structure normal, the radiation is assumed to be incident in
$(xz)$ plane, $\hat{\bm e}$ is the unit vector in light
propagation direction and $\bm e$ is the (complex) polarization
vector of radiation, $P_{\rm circ}$ is the circular polarization
degree and $\bm q$ is the radiation wave vector.
The  {dependence of the photocurrent components on}
 the radiation polarization state, incidence angle and frequency  {fully agrees} with  {the theory results} in~\cite{karch2010,2010arXiv1002.1047K,GlazovGanichev}. Moreover,  {only assuming short-range scattering,} the microscopic theory yields the absolute value of the photocurrent without fitting parameters~\cite{karch2010}.  {Since} conduction  and valence band are symmetric with respect to the Dirac point, the signal reverses its sign by changing  {from $p$ to $n$-type carriers}.

\section{Photogalvanics and reststrahl band 
assisted photocurrents in epitaxial graphene layers}
\label{pge}

\begin{figure}
  \includegraphics[width=\columnwidth]{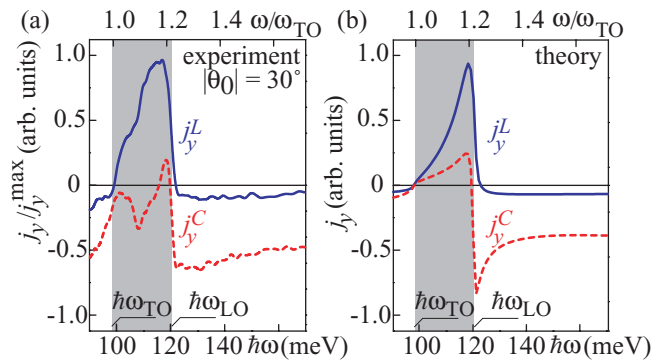}%
  \caption{\label{figure4restsrtahlen}
    Spectral behaviour of the linear (solid lines) 
and circular (dashed lines) photocurrents excited by radiation 
in the frequency range of reststrahlen band of SiC substrate indicated by a grey background.
(a)  Experimental results. 
(b)  Calculated photocurrents using 
a ratio of the photon drag to photogalvanic effects equal to -0.3.
Data are given after
 \cite{reststr2013}.
}
\end{figure}

 {To break inversion symmetry, necessary to observe photogalvanic effects (PGE)~\cite{3authors}, flat infinite graphene layers can be placed on a substrate or host adatoms on one surface only.}
This structure inversion asymmetry removes the  {$z$ and $-z$ equivalence} 
and reduces the symmetry to the $C_{6\rm v}$ point group. The photogalvanic effects give rise to the
linear and circular photocurrents~\cite{2010arXiv1002.1047K}:
\begin{subequations}
\label{j:pge}
\begin{equation}
\label{j:pge:x}
j_{x}/E^2 = \chi_{l} \frac{e_x e_z^*+ e_x^*e_z}{2} \:,
\end{equation}
\begin{equation}
\label{j:pge:y}
j_{y}/E^2 = \chi_{l} \frac{e_y e_z^*+ e_y^*e_z}{2} +
\chi_{c} P_{\rm circ} \hat{ e}_x  \:,
\end{equation}
\end{subequations}
described by two independent parameters $\chi_l$ and $\chi_c$.
 {Unlike in conventional semiconductor quantums well or heterostructures, where the wavefunctions spread over many atomic layer, in graphene carriers are strongly confined to strictly two dimensions and therefore almost do not react to an electric field in $z$-direction. Consequently, the PGE in graphene is reduced when a $z$-component of the radiation field is present.}
 {Since both PGE and photon drag show a similar response to polarization, the stronger drag effect usually masks the PGE.}
 {To observe the PGE more clearly, the photon drag contribution needs to be reduced, for instance by using high radiation frequencies.}

 {Similar to the orbital mechanisms of the PGE in conventional semiconductor nanostructures, the THz induced PGE here is caused by the quantum interference of the Drude-like indirect optical transitions}~\cite{Tarasenko2007,PhysRevB.79.121302,tarasenko11}. 
 {We observed both linear and circular PGEs in epitaxial graphene samples using mid-infrared radiation of about 30~THz.}
 {The observation of PGE is facilitated by the suppression of the photon drag effect at high frequencies and also by the fact that photogalvanic and drag effects lead to opposite signs in their respective contributions to photocurrent. Therefore, by varying the radiation frequency a sign change in the photocurrent was observed, confirming the existence of a PGE with substantial amplitude~\cite{jiangPRB2010}.}
 {The photocurrent due to the circular PGE closely matches the value obtained by a theoretical estimate for a sufficiently strong degree of asymmetry, $\langle V_0V_1\rangle/{\langle V_0^2\rangle} \approx 0.5$. }
 {We stress that the PGE requires structure inversion asymmetry, which is not present in graphene where the $z\rightarrow -z$ symmetry is preserved, for example, clean, free standing graphene.}
 {As will be detailed in Sec.~\ref{magnet}, the observation of the magnetic quantum ratchet effect in epitaxial graphene constitutes a nice example
of a large structure inversion asymmetry~\cite{Drexler2013} caused by adatoms or the substrate.}
Our studies revealed a further  {interesting feature of 
photoelectric effects in graphene: A 
resonance-like 
frequency} dependence for frequencies lying within the reststrahl band of the SiC substrate~\cite{reststr2013}, see Fig.~\ref{figure4restsrtahlen}(a). In particular, photocurrents excited by linearly polarized radiation
are strongly enhanced just in the range of reststrahlen band, i.e. for frequencies 
at which the reflection coefficient is close to 100~\%. 
 {The photocurrent consists of photon drag and PGE contributions of similar strength, responding to the in-plane and out-of-plane components of the local electric field felt by the electrons in the graphene layer. The field distribution at a distance $d \approx$ 2~\AA \, from the SiC surface (the position of the graphene layer in our samples, see Ref.~\cite{Borysiuk}) can be
calculated using the macroscopic Fresnel formulas. This model describes the observed resonance surprisingly well.}
The result of the corresponding photocurrent calculations is shown in 
Fig.~\ref{figure4restsrtahlen}(b).
Importantly, those observations demonstrate that by engineering the substrate material and spectral range, we can greatly enhance non-linear optical and opto-electronic effects in 2D materials.

\section{Edge photocurrents}
\label{edge}

\begin{figure*}
  \includegraphics[width=0.7\linewidth]{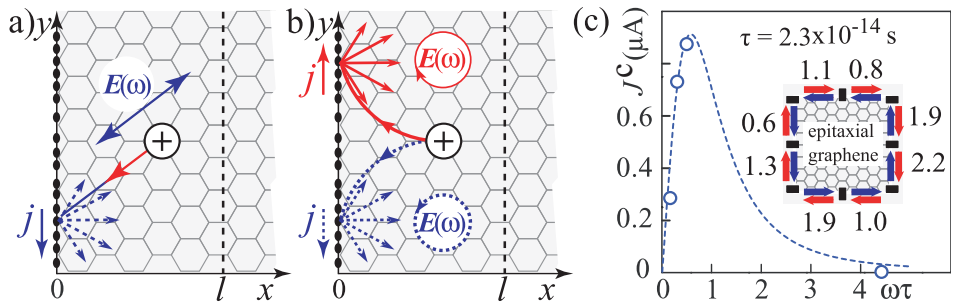}%
  \caption{\label{figure5edge} 
	 {(a) Sketch of the generation of edge photocurrents under illumination by linear polarized radiation (field $\bm E(\omega)$, blue double arrow). Charge carriers (holes in this case) follow the external field. In one half-cycle of the radiation, they are accelerated towards the sample edge (red arrow), where they are scattered and lose their momentum memory (dashed blue arrows), leading to an edge current $\bm{j}$ within a mean free path $\ell$ from the edge.
(b) For circularly polarized radiation, carriers move on circular orbits, whose sense of rotation depends on the radiation helicity. Similar to (a) they are scattered at the sample edges and generate a net current.
(c) Frequency dependence of the circular component of the edge current showing a maximum at $\omega \tau = 1$. Inset: Magnitude and direction of the circular edge current for opposite helicities (red and blue arrows) showing that the same direction of rotation is maintained along the entire sample boundary.}
 Data are given after  
 \cite{edgePRL2011}.
}
\end{figure*}

According to the symmetry analysis given in Sec.~\ref{symmetry}
illumination of pristine graphene by radiation at normal incidence 
does not cause an electric current. 
 {When the sample edges are illuminated, however, inversion symmetry is broken, and edge photocurrents can be observed.}
In  Fig.~\ref{figure5edge}(a) we illustrate the microscopic process actuating the edge photocurrent. Linearly polarized radiation acts on the free carriers in the semi-infinite graphene plane ($x>0$).  For $\omega \tau<1$, the drift motion of the carriers follows the radiation electric field. In one half of the radiation cycle, carriers are moving away from the sample edge. In the other half cycle, they are accelerated towards the sample edge and eventually scattered by edge roughness, randomizing their momentum. On average, they perform a directed motion which is dependent on the angle between linear polarization and the sample edge, resulting in the linear photogalvanic effect. 
 {In a narrow stripe close to the sample edge, up to a distance of roughly the mean free path $\ell$, an electric current is generated.}
Under circularly polarized radiation, curved trajectories emerge (see Fig.~\ref{figure5edge}(b)), as the carriers try to follow the external electric field. This results in a current  {reversing its sign, when the radiation helicity changes from $\sigma^+$ (solid) to $\sigma^-$ (dashed).}
 {For fixed helicity irradiation of opposite sample edges also results in the opposite sign of the photocurrent.}

 {Boltzmann kinetic equation for $f(\bm{p},x,t)$ describing the edge photocurrents is given by:}
\begin{equation}\label{f_generaledge}
\frac{\partial f}{\partial t} + v_x  \frac{\partial f}{\partial x}
+ q \bm{E}(t) \frac{\partial f}{\partial \bm{p}} = Q\{f\} \:,
\end{equation}
 {where coordinate $x \geq 0$ for a semi-infinite layer,} 
$q$ is the carrier charge ($q= + |e|$ for holes and $- |e|$ for
electrons), and $Q\{f\}$ is the collision integral.
The distribution function can be expanded in series of powers of the electric field,
\begin{equation}
\label{f:genseriesedge}
f(\bm{p},x,t) = f_0(\varepsilon_{\bm{p}}) + [f_1 (\bm{p},x){\rm e}^{- \mathrm i \omega t} + {\rm c.c.}]
+ f_2(\bm{p},x) + ... \:,
\end{equation}
where $f_0(\varepsilon_{\bm{p}})$ is the equilibrium distribution
function,
$f_1 \propto |\bm{E}|$, and $f_2 \propto |\bm{E}|^2$.
The  {oscillating with frequency $\omega$ first order in electric field $\bm E$ correction $f_1 \propto |\bm{E}|$
does not contribute to a $dc$ current.}
 {Thus, }the $dc$ current along the structure edge is  {due to}
the \textit{second order}  \textbf{\textit{E}}-field correction $f_2$ and
given by 
\begin{equation}\label{J_edge_current}
J_y = 4\ q \int_{0}^{\infty} dx \sum_{\bm{p}} f_2(\bm{p},x) v_y \:.
\end{equation}
 {Here factor~4 takes into account the spin and valley degeneracy.}
The analysis  {shows}
that the total current consists of several contributions proportional to 
four Stokes parameters, which  {all} are observed in experiment.

While the edge photocurrents are detected in both epitaxial and in exfoliated samples, 
 {in large-area graphene the analysis of the experiments is substantially easier. 
In  exfoliated graphene, opposite edges of $\mu$m-sized flakes are illuminated inevitably. 
In contrast in large-area} samples only a single edge can be illuminated. In particular, scanning the laser beam across the 
sample edges, demonstrated that the  {coordinate dependence of the signal almost reproduce the}  
the Gaussian  {beam} profile.
 {The red and blue arrows in the inset in Fig.~\ref{figure5edge}(c) illustrate
the current directions for  $\sigma^+$ and  $\sigma^-$ circularly polarized radiation 
and the numbers show the magnitude of the circular photocurrent $J_A$ for various contact pairs}
In these measurements the 
laser spot is  always {placed} 
between the contacts at which the signal is picked-up, preventing a temperature gradient between contacts. 
Remarkably,  the  edge photocurrent 
 {proceeds in the same sense of rotation along the edges of the square shaped samples and changes its direction when reversing from $\sigma^+$ to  $\sigma^-$polarization.}

Edge photocurrents have been detected in a wide range of radiation frequencies.
Figure~\ref{figure5edge}(c)
 shows the circular edge photocurrent $J_A \propto \sin{2 \varphi}$ excited by THz radiation 
as a function of $\omega \tau$, where $\tau$ is the scattering time at a sample edge.
 {The frequency dependency and magnitudes of the circular edge current agrees well with theory. The only 
parameter used for fit in Fig.~\ref{figure5edge}(c)} is the scattering time  {close to} the edge,  {which was found to be quite close to the average bulk scattering time.}
The small differences can most probably be explained by inhomogeneities in the distribution of scatterers
Moreover, the sign of the  {photocurrent excited at fixed helicity}
reflects the type of the charge carriers  {close to }the edge.   {The latter has been shown
to be holes even} for $n$-type epitaxial graphene. 
 {This is in accordance with scanning Raman experiments pointing to a $p$-type doping at}
graphene 
edges \cite{Raman1,Raman3},
transport measurements,  {where a} transition from $n$-to $p$-type 
at the edges of graphene flakes on SiO$_2$  {is reported}~\cite{SK_TS4}
and growth details of epitaxial graphene~\cite{Emtsev2009,Tzalenchuk2010,erl5}.
 {Thus, edge photocurrents may be used to characterize graphene edge properties up to room temperature.}

\section{ Magnetic quantum ratchet effect}
\label{magnet}

\begin{figure}
  \includegraphics[width=\columnwidth]{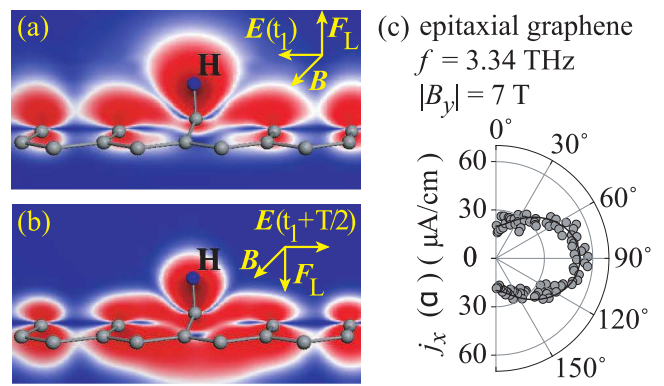}%
  \caption{\label{figure6magnetic}
  (a,b) Electron density distribution in graphene with a hydrogen adatom for two moments in time separated by half a radiation period.
	 {(c) Angular dependence of the ratchet current. $\alpha$ denotes the angle between the external magnetic field and the radiation electric field. Dots: experimental data taken at $T=115$ K, $B=7$~T and field amplitude of 10 kV/cm. Solid line: theory.}
 Data are given after 
\cite{Drexler2013}.
		}
\end{figure}

The magnetic quantum ratchet effect has been observed in 
single-layer graphene samples excited with a pulsed molecular terahertz laser and  
subjected to an in-plane magnetic field. 
The physics behind the magnetic quantum ratchet effect is illustrated in Figs.~\ref{figure6magnetic}(a) and (b). 
 {Dirac electrons are driven by the time-dependent electric field $\bm{E}(t)$ and move in alternating directions in the graphene plane. The external static magnetic field $\bm{B}$ leads to a Lorentz force, deflecting the right-moving electrons upwards and the left-moving electrons downwards }%
(see Fig.~\ref{figure6magnetic} for an illustration at times $t_1$ and 
$t_2=t_1+T/2$ differing by a half a period $T$ of the
radiation electric field). 
 {For spatially symmetric systems this would lead to a zero $dc$ current. However, when, e.g., top adsorbates are present, spatial symmetry is broken and } 
electrons shifted above or below the graphene plane experience different degrees of disorder, which results in a non-zero $dc$ current. 
We note that  {a similar mechanism was discussed for }
inversion channels in Si and semiconductor 
quantum wells~\cite{Falko89,Tarasenko08,Tarasenko11}.

 {The current is proportional to the}  {square of $ac$ electric field amplitude and  the magnetic field strength.
Reversing the direction of static magnetic field changes the sign of the photocurrent. It also depends on the 
angle $\alpha$ between the $ac$ electric field $\bm{E}(t)$ and the static magnetic field $\bm{B}$. 
A characteristic polarization dependence of the magnetic 
ratchet current is shown in Fig.~\ref{figure6magnetic}(c). The current is maximal} 
 { for perpendicular electric and magnetic fields and remaining non-zero for parallel fields.}%
 {The current is well fitted }
by the equation
 $j_x = j_1 \cos 2 \alpha + j_2$ with two contributions $j_1$ and $j_2$.
 {Exactly this behaviour follows from the phenomenological and microscopic theory developed
in Ref.~\cite{Drexler2013} and described below.}

{As an important}  {fact, when comparing two kinds of graphene samples with different surface treatment, we find opposite signs of the slope $j_x(|B_y|)$.} While surfaces of samples with graphene encapsulated in a thin polymer film exhibit a positive slope
photocurrent, in the sample with unprotected surface the slope is negative,  {proving differents signs of structure inversion asymmetry (SIA) for both kinds of samples.}

{We developed a microscopic theory of the observed effect, 
which agrees beautifully with the experiments and is supported by first-principles calculations.}
The electric current density is calculated using the general expression
Eq.~(\ref{generalcurrent})
\begin{equation}\label{current_gen_magnetic}
\bm{j} = 4 e \sum_{\bm{p}} \bm{v} f(\bm{p},t) \:.
\end{equation}
The distribution function can be  {obtained} from the Boltzmann equation
\begin{equation}\label{kineqmagnetic}
\frac{\partial f(\bm{p},t)}{\partial t} + e \bm{E}(t) \cdot
\frac{\partial f(\bm{p},t)}{\partial \bm{p}} =       Q\{f\}\:.
\end{equation}
For elastic scattering, it has the form
\begin{equation}\label{Stmagnetic}
Q\{f\} = \frac{2\pi}{\hbar} \sum_{\bm{p}'}
\langle |V_{\bm{p}'\bm{p}}|^2 \rangle [f(\bm{p}',t) - f(\bm{p},t)] \,
\delta(\varepsilon - \varepsilon') \:,
\end{equation}
where the angular brackets denote impurity ensemble averaging and
$V_{\bm{p}'\bm{p}}$ is the matrix element of electron scattering
between the initial and final states with the momenta $\bm{p}$ and $\bm{p}'$,
respectively. The ratchet currents  {originate in } the asymmetry of electron scattering,
which is caused by the $\sigma\,-\,\pi$ hybridization around the Dirac
points in the in-plane magnetic field.
Formally, for the magnetic field $B_y$, it is described by the matrix element
\begin{equation}\label{V_ppmagnetic}
V_{\bm{p}'\bm{p}} = V_{\pi\pi} - B_y(p_x+p'_x) \frac{z_{\pi\sigma} e}{\varepsilon_{\pi\sigma} m_0 c } V_{\pi\sigma}  \:,
\end{equation}
where $z_{\pi\sigma}$ is the coordinate matrix element between the $\pi$- and
$\sigma$-band states, $\varepsilon_{\pi\sigma}$ is the energy distance between
the two bands, $V_{\pi\pi}$ and $V_{\pi\sigma}$ are the intraband and interband
matrix elements of scattering at zero magnetic field,
$m_0$ is the free electron mass, and $c$ is the speed of light.

After solving the Boltzmann equation, the analysis of the photocurrent as a function of the radiation
polarization reveals that the total current consists of several 
contributions proportional to four Stokes parameter. Individual 
contributions proportional to the Stokes parameters describing the 
degree of linear polarization and 
the one given by the radiation helicity describes the 
linear and circular magnetic quantum ratchet photocurrents.
{Experiments applying radiation with different polarization states demonstrate that all these photocurrents can  be excited efficiently in graphene.}

\section{Terahertz ratchet effects in graphene with a lateral superlattice}
\label{ratchet}

\begin{figure}
  \includegraphics[width=\columnwidth]{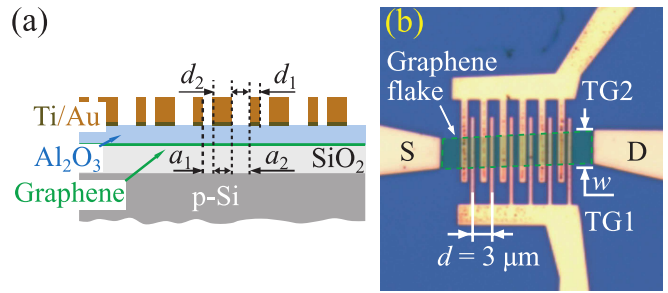}%
  \caption{\label{figure7ratchet1}
Cross-section (a)  and an optical micrograph (b) of the interdigitated grating-gates: 
The supercell of the grating gate fingers consists of metal stripes
having two different widths $d_1 = 0.5~\mu$m and $d_2 = 1$~$\mu$m
separated by spacings $a_1 = 0.5$~$\mu$m and $a_2 = 1$~$\mu$m. 
This asymmetric supercell is  repeated six times to create a periodic asymmetric potential 
with period $d = d_1+ d_2+a_1+a_2 = 3$~$\mu$m, see panel (b). Data are given after~
 \cite{ratchet_graphene16}.
}
\end{figure}

\begin{figure}
  \includegraphics[width=\columnwidth]{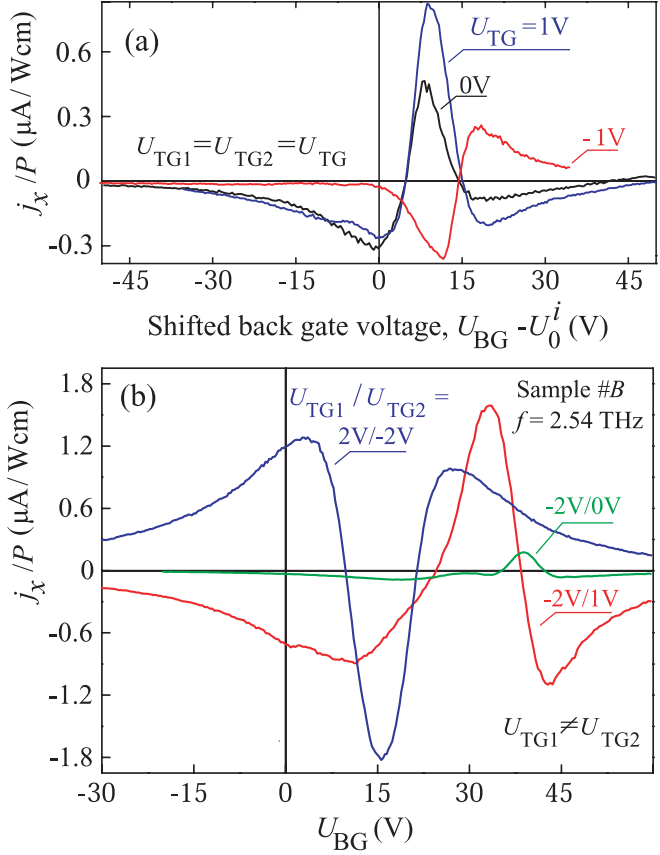}%
  \caption{\label{figure8ratchet2}
  (a) Photocurrent $j_x(\alpha = 0)$ normalized by the radiation power as a function of the relative gate voltage $U_{\rm BG} - U^i_0$, 
where $U^i_0$ is defined as the back-gate voltage for which the resistance is the largest at corresponding $U_{\rm TG}$, see panel (a). 
(b) Gate voltage dependence of the photocurrent $j_x(\alpha = 0)$, $U_{\rm TG1}\neq U_{\rm TG2}$. Insets show carrier density and energy band offset profiles at $U_{\rm BG}=-20\thinspace\mathrm{V}$. Data are given after 
\cite{ratchet_graphene16}.
}
\end{figure}

Ratchet effects discussed in the previous section require a static magnetic field.
Another efficient way to generate a $dc$ electric current caused by ratchet effect 
implies symmetry reduction due to deposition of a periodic asymmetric lateral
metal structure on the top of graphene. This type of graphene ratchets
has been experimental realized and systematic study 
in both (i) epitaxially grown and (ii) exfoliated graphene with an asymmetric lateral periodic potential~\cite{ratchet_graphene16}. The modulated potential has been obtained by  fabrication of either a sequence of metal stripes  on top of graphene or
 inter-digitated comb-like dual-grating-gate structures. The latter structure is shown in Figs.~\ref{figure7ratchet1}(a) and (b). Our work demonstrated that 
 {a polarization dependent $dc$ current can be generated by exposing a modulated device to THz laser radiation.}
By applying different voltages to the two gratings, we can control the photocurrent behaviour at different structure asymmetry, carrier type and density. A typical behaviour of the ratchet photocurrents upon variation of back gate potential is shown in Fig.~\ref{figure8ratchet2}(a) and (b) for various combination of the top dual-grating-gate potentials. Figure~\ref{figure8ratchet2}(a) shows the data for equipotential top gates as a function of the effective back gate which is defined as $U_{\rm BG} - U_0^i$, where $U_{\rm BG}$ is the applied back gate voltage and $U_0^i$ {are  back gate voltages of} the charge neutrality point {measured for the corresponding top gates voltages}. It is seen that the signals are strongly enhanced in the vicinity of the Dirac point and have opposite sign for opposite top gate voltages. A substantial signal is also obtained for zero top gate voltage. This signal is due to the electrostatic potential caused by metal film placed in
the proximity of graphene. The data presented in Fig.~\ref{figure8ratchet2}(b)
 reveal that the photocurrent reflects the degree of asymmetry induced by different top gate potentials and even vanishes for a symmetric profile. 
The  measurements together with a beam scan across the lateral structure 
prove that the observed photocurrent stems from the ratchet effect.

The experimental data and the theoretical model are discussed by taking 
the calculated potential profile and near-field effects explicitly into account.  
The ratchet current consists of the
Seebeck thermo\-ratchet effect as well as the ``linear'' and ``circular'' ratchets, sensitive to the corresponding polarization of the driving electromagnetic force.
The results are analyzed in terms of electronic and  plasmonic mechanisms of a photocurrent in periodic structures.
The ratchet photocurrent appears due to the noncentrosymmetry of the periodic graphene structure unit cell. 
 The effect of the grating is twofold: (i)~it generates a  one-dimensional periodic electrostatic potential ${\cal V}(x)$ acting upon the 2D carriers and (ii)~it causes a spatial modulation of the THz electric field due to the near field diffraction\,\cite{ratchet_graphene16,ratchet2009,ratchet2011,Review_JETP_Lett,Theory_PRB_11}.
These one-dimensional asymmetries result in the generation of a $dc$ electric current. The ratchet current may flow perpendicular to the metal fingers or along them. The mechanism leading to the photocurrent formation can be illustrated on the basis of the photocurrent caused by the Seebeck ratchet effect (thermoratchet). 
This type of ratchet currents can be generated in the direction perpendicular to the metal stripes and corresponds to the polarization independent photocurrent.

The spatially-modulated  electric field of the radiation heats the electron gas to $T(x) = \bar{T} + \delta T(x)$~\cite{footnoteLG}.
Here $\bar{T}$ is the average electron temperature and 
$\delta T(x)$ oscillates along the $x$-direction with the superlattice period 
$d$.
 In turn, the nonequilibrium correction $\delta T(x)$ causes an inhomogeneous
correction to the $dc$ conductivity,
$\delta \sigma(x) \propto \delta T(x)$. 
Taking into account the spatially modulated electric field $(-1/e) d{\cal V}/dx$ we obtain from Ohm's law 
the thermoratchet current~\cite{Theory_PRB_11}
\begin{equation} \label{j_Seeb}
j_x^S =- {1 \over e} \left \langle {d {\cal V} \over dx} \delta \sigma(x) \right\rangle \:.
\end{equation}
Here $e<0$ is the electron charge, and angular brackets denote averaging over a spatial period. This photocurrent vanishes if the temperature is spatially uniform, therefore it is called the Seebeck ratchet current~\cite{footnoteT1}.

Besides the thermoratchet effect the THz radiation can induce additional photocurrents being
sensitive to the linear polarization plane orientation or to the helicity of 
circularly polarized photoexcitation.   {These photocurrents have been observed in epitaxially grown
graphene with lateral superlattice~\cite{ratchet_graphene16}. Apart a novel experimental access to the light matter interaction in graphene these results may also have an application potential. 
Photon helicity driven ratchet current can be utilized for a novel kind of all-electric ellipticity meter.  
So far such devices implements circular photogalvanic effect in semiconductor 
quantum wells~\cite{ellipticitydetector,ellipticitydetector2}. Unique 
nonlinear properties of graphene together with the advantages of ratchet photoresponse~\cite{Review_JETP_Lett,16w,31,115,otsuji2,Otsuji_Ganichev} can 
substantially improve detectors detectivity and time resolution as well as extend the operation spectral range. 
}

 {To summarize this part,} experiments on two different types of graphene structures provided 
a self-consistent picture demonstrating that the photocurrents 
 (i) are generated due to the presence of asymmetric superlattices, (ii) are characterized by specific polarization dependencies for directions along and across the metal stripes, (iii) change direction upon reversing the in-plane asymmetry of the electrostatic potential as well as  changing the carrier type,
(iv) are characterized by a complex sign-alternating back gate 
voltage dependence in the vicinity of the Dirac point, (v) are strongly enhanced around the Dirac point
and  {(vi) have potential for development of the all-electric ellipticity meter}.

\section{Fast room temperature detectors of THz radiation}
\label{detector}

\begin{figure}
  \includegraphics[width=\columnwidth]{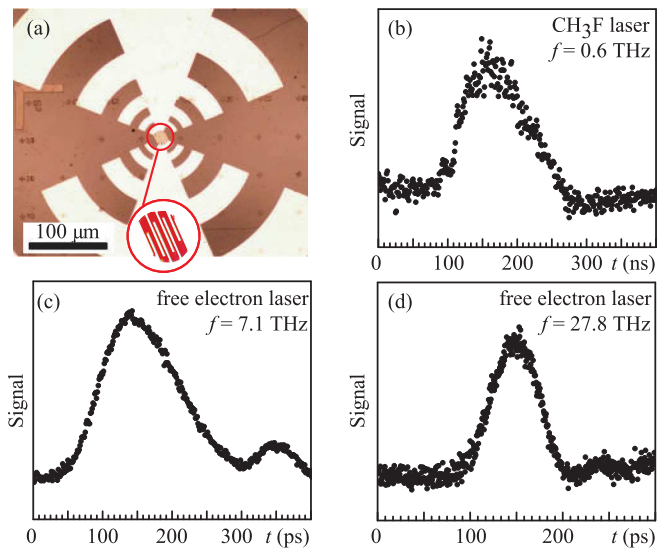}%
  \caption{\label{figure9detector}
    (a) 
		 {Micrograph of the 
		antenna with the graphene flake below the inter-digitated electrodes.} 
		Fast response of the detectors at different frequencies obtained with two types of lasers.  (b) 0.6~THz , (c) 7.1~THz, (d) 27.8~THz. 
		Data are given after 
		\cite{detectorFEL}.
		}
\end{figure}

Finally, studying THz radiation induced opto-\-electronic phenomena in 
graphene is of particular importance not only for for exploring the physical properties of these materials but also for the development of novel THz radiation detectors. 
In this section we describe an ultrafast bolometric room temperature graphene based THz detector showing 40~picosecond electrical rise time over a spectral range that spans nearly three orders of magnitude, from the visible to the far-infrared\cite{detectorFEL1,detectorFEL}. The detector employs a graphene active region with inter-digitated electrodes that are connected to a log-periodic antenna to improve the long-wavelength collection efficiency, see Fig~\ref{figure9detector}(a), and a silicon carbide substrate that is transparent throughout the visible regime. The detector exhibits a noise-equivalent power of approximately {100~$\mu$W $\cdot$ Hz$^{-1/2}$} and is characterized at frequencies from 0.6 to 384~THz (wavelengths from 500 $\mu$m
to 780 nm). To estimate the noise-equivalent power a calibrated photon-drag detector was used~\cite{Ganichev84p20}.

Low frequency measurements have been performed with a pulsed THz CH$_3$F laser operating at a frequency of 0.6 THz~\cite{tunnelreview02}. 
 {The laser pulses (about 200 ns duration) are composed of many short peaks (about 1 ns).
The repetition rate is 1 Hz and the signals are recorded with a standard digital oscilloscope.}

Figure~\ref{figure9detector}(b) shows the graphene detector signal. While the rise time of the detector is 
 {less than}
the pulse duration the detector can be used to analyze the pulse shape. Similar results have been obtained at higher frequencies ranging from 1 to 5 THz by using NH$_3$ as laser active medium. 

To explore the time resolution of the graphene detector and extend frequency range to higher frequencies measurements at the free-electron laser FELBE (Dresden-Rossendorf) were performed. The laser provides a pulse train with a repetition rate of 13 MHz at frequencies between 1.3   and 60 THz.
Pulse traces for frequencies 7.1 and~27.8~THz are shown in Figs.~\ref{figure9detector}(c) and (d), respectively. The rise time is mainly limited by the parasitic capacitance of the antenna and the inductance of the electrical connections. 
 {The intrinsic response time of graphene was determined to be about 10 ps using optical autocorrelation measurements~\cite{Cai,detectorFEL1}.}

 {To our knowledge, a similar broad and continuous frequency range of a fast detector was not reported before, and is unique to our device, where 
graphene as a detector material and SiC substrates are combined. Also, we did not find a significant change in detector response when changing the excitation frequency from inside to just outside the reststrahlen band.}
The pulse trace shown in Fig.~\ref{figure9detector}(d) is obtained for frequency $f=27.8$~THz 
lying within the reststrahlen band. 

 {Our graphene-bases detector enables ultrafast room-temperature detection in a broad frequency range. Given the extremely low heat capacity of charge carriers in graphene, which are heated directly by the incoming radiation, the electron temperature responds strongly. On the other hand, electrons can cool fast and efficiently via optical phonons~\cite{Hartmann}.}
The presented detector is well suited for a  {great variety} of pulsed laser sources like optical-parametric oscillators and amplifiers or difference-frequency mixers, which makes it a very  {promising} device for multicolor ultrafast spectroscopy.

\section{Conclusions and outlook}
\label{summary}

 {The physics of nonlinear electron transport and optical phenomena in graphene has already resulted in a great variety of fascinating effects. We still need to develop a full understanding of many of the effects using new experimental and theoretical concepts. For example, tuning the non-linear response using external magnetic fields, strain or by combining graphene with other 2D materials will lead to new insights.  
Moving beyond graphene, we expect that similar effects can be studied in boron nitride, transition metal dichalcogenieds and topological insulators. For the latter, first experiments have been performed recently~\cite{HosurBerry,TopIns4,Ultrathin,TopIns2,TopIns5,TItheory,TopIns3,Ultrafast,TopIns1,TopIns2add}. Finally, from an application point of view, we believe that the described effects will come in useful for material characterization as well as new non-linear devices based on graphene.
}

\section*{Acknowledgements}

{We are grateful to M.M. Glazov for reading the manuscript and valuable discussions.}
Financial supported of the  DFG  (SPP~1459 and GRK~1570) is gratefully acknowledged.


\begin{thebibliography}{160}
\expandafter\ifx\csname natexlab\endcsname\relax\def\natexlab#1{#1}\fi
\expandafter\ifx\csname bibnamefont\endcsname\relax
  \def\bibnamefont#1{#1}\fi
\expandafter\ifx\csname bibfnamefont\endcsname\relax
  \def\bibfnamefont#1{#1}\fi
\expandafter\ifx\csname citenamefont\endcsname\relax
  \def\citenamefont#1{#1}\fi
\expandafter\ifx\csname url\endcsname\relax
  \def\url#1{\texttt{#1}}\fi
\expandafter\ifx\csname urlprefix\endcsname\relax\def\urlprefix{URL }\fi
\providecommand{\bibinfo}[2]{#2}
\providecommand{\eprint}[2][]{\url{#2}}


\bibitem{Neto}  A.H. Castro Neto, F. Guinea, N.M.R. Peres, K.S. Novoselov, A.K. Geim, 
Rev. Modern Phys. \textbf{81}, 109 (2009).

\bibitem{Peres}  N.M.R. Peres, 
Rev. Modern Phys. \textbf{82},  2673 (2010).

\bibitem{Avouris}P. Avouris, 
Nano Lett. {\bf 10}, 4285 (2010).

\bibitem{Sarma}  S. Das Sarma, S. Adam, E.H. Hwang, E. Rossi, 
Rev. Modern Phys. \textbf{83}, 407 (2011).

\bibitem{Young} A. F. Young and P. Kim,
Ann. Rev.  Cond. Mat. Phys. {\bf 2}, 101 (2011).

\bibitem{McCann}  E. McCann, M. Koshino, 
Rep. Prog. Phys. \textbf{76}, 056503 (2013).



\bibitem{Bonaccorso}  F. Bonaccorso, Z. Sun, T. Hasan, A.C. Ferrari, 
Nature Photonics \textbf{4}, 611 (2010).

\bibitem{mueller} T. Mueller, F. Xia, and P. Avouris,
Nature Photon. {\bf 4}, 297 (2010).

\bibitem{novoselov} T.J. Echtermeyer, L. Britnell, P.K. Jasnos, A. Lombardo, R.V. Gorbachev, A.N. Grigorenko, A.K. Geim, A.C. Ferrari, and K.S. Novoselov,
Nature Commun. {\bf 2}, 458 (2011).

\bibitem{GlazovGanichev} M.M. Glazov and S.D. Ganichev,
Physics Reports \textbf{535}, 101 (2014) (2014).

\bibitem{karch2010} J.~Karch, P.~Olbrich, M.~Schmalzbauer, C.~Zoth, C.~Brinsteiner,   M.~Fehrenbacher, U.~Wurstbauer, M.~M. Glazov, S.~A. Tarasenko, E.~L.   Ivchenko, D.~Weiss, J.~Eroms, R.~Yakimova, S.~Lara-Avila, S.~Kubatkin, S.~D.
Ganichev, 
Phys. Rev. Lett. \textbf{105}, 227402 (2010).

\bibitem{2010arXiv1002.1047K}
J.~{Karch}, P.~{Olbrich}, M.~{Schmalzbauer}, C.~{Brinsteiner}, U.~{Wurstbauer},
  M.~M. {Glazov}, S.~A. {Tarasenko}, E.~L. {Ivchenko}, D.~{Weiss}, J.~{Eroms},
  S.~D. {Ganichev}, 
arXiv cond-mat 1002.1047  (2010).


\bibitem{jiangPRB2010}
C.~Jiang, V.~A. Shalygin, V.~Y. Panevin, S.~N. Danilov, M.~M. Glazov,
  R.~Yakimova, S.~Lara-Avila, S.~Kubatkin, S.~D. Ganichev, 
Phys. Rev. B \textbf{84}, 125429 (2011).
	
	
\bibitem{reststr2013} P. Olbrich, C. Drexler, L. E. Golub, S. N. Danilov, V. A. Shalygin, 
R. Yakimova, S. Lara-Avila, S. Kubatkin, B. Redlich, R. Huber, and S. D. Ganichev,
Phys. Rev. B \textbf{88}, 245425 (2013).


\bibitem{edgePRL2011}
J.~Karch, C.~Drexler, P.~Olbrich, M.~Fehrenbacher, M.~Hirmer, M.~M. Glazov,
  S.~A. Tarasenko, E.~L. Ivchenko, B.~Birkner, J.~Eroms, D.~Weiss, R.~Yakimova,
  S.~Lara-Avila, S.~Kubatkin, M.~Ostler, T.~Seyller, S.~D. Ganichev,
Phys. Rev. Lett. \textbf{107}, 276601 (2011).

\bibitem{sun2010}
D.~Sun, C.~Divin, J.~Rioux, J.~E. Sipe, C.~Berger, W.~A. de~Heer, P.~N. First,
  T.~B. Norris.
\newblock Nano Lett. \textbf{10}, 1293 (2010).



\bibitem{Sun:2012ys}
D.~Sun, J.~Rioux, J.~E. Sipe, Y.~Zou, M.~T. Mihnev, C.~Berger, W.~A. de~Heer,
  P.~N. First, and T.~B. Norris,
Phys. Rev. B {\bf 85}, 165427 (2012).

  
  \bibitem{2012winzent}
D.~Sun, C.~Divin, M.~Mihnev, T.~Winzer, E.~Malic, A.~Knorr, J.~E. Sipe,
  C.~Berger, W.~A. de~Heer, P.~N. First, and T.~B. Norris,
{New Journal of Physics} {\bf 14}, 105012 (2012).

\bibitem{Drexler2013} C. Drexler,
S. A. Tarasenko, P. Olbrich, J. Karch, M. Hirmer, F. Muller, M. Gmitra, J. Fabian, R. Yakimova, S. Lara-Avila, S. Kubatkin, M. Wang, R. Vajtai, P.M. Ajayan, J. Kono, and S. D. Ganichev, 
{Nature Nanotechnology} {\bf 8}, 104 (2013).


\bibitem{ratchet_graphene16} P. Olbrich, J. Kamann, M. K\"{o}nig, J. Munzert, L. Tutsch, J. Eroms, D. Weiss, M.-H. Liu, L. E. Golub, E. L. Ivchenko, V. V. Popov, D. V. Fateev, K. V. Mashinsky, F. Fromm, Th. Seyller, and S. D. Ganichev, 
Phys. Rev. B \textbf{93}, 075422 (2016). 

\bibitem{Prechtel:2012kx}
L.~Prechtel, L.~Song, D.~Schuh, P.~Ajayan, W.~Wegscheider, and A.~W.
Holleitner, 
{Nature Communications} {\bf 3}, 01 (2012).

\bibitem{Graham:2013uq} M.~W. Graham, S.-F. Shi, D.~C. Ralph, J.~Park, and P.~L. McEuen, 
\newblock {Nature Physics} {\bf 9}, 103 (2013).

\bibitem{detectorFEL1}  M. Mittendorff, S. Winnerl, J. Kamann, J. Eroms, D. Weiss, H. Schneider, and M. Helm,
Appl. Phys. Lett. {\bf 103}, 021113 (2013).

\bibitem{detectorFEL}  M. Mittendorff, J. Kamann, J. Eroms, D. Weiss, C. Drexler, S.D. Ganichev, J. Kerbusch, A. Erbe, R.J. Suess, T.E. Murphy, S. Chatterjee, K. Kolata, J. Ohser, J.C. Koenig-Otto, H. Schneider, M. Helm, and S. Winnerl, 
Optics Express \textbf{23}, 
28728
(2015).



\bibitem{plasmawave} L. Vicarelli, M. S. Vitiello, D. Coquillat, A. Lombardo, A. C. Ferrari, W. Knap, M. Polini, V. Pellegrini, and A. Tredicucci,
Nature Materials {\bf 11}, 865 (2012).


\bibitem{Freitag} M. Freitag, T. Low, F. Xia, and P. Avouris, 
Nat. Photonics 7, 53 (2013).

\bibitem{koppens} F. H. L. Koppens, T. Mueller, Ph. Avouris, A. C. Ferrari, M. S. Vitiello, and M. Polini,
Nature Nanotech. {\bf 9}, 780-793 (2014).

\bibitem{Hartmann}	R. R. Hartmann, J. Kono, and M. E. Portnoi, 
Nanotechnology \textbf{25}, 322001 (2014).

\bibitem{Tredicucci14} A. Tredicucci and M.S. Vitiello, 
IEEE J. Sel. Top. Quant. Electr.  \textbf{20}, 8500109  (2014).





\bibitem{sturmanBOOK}
B.~Sturman, V.~Fridkin.
{\textit{The photovoltaic and photorefractive effects in
  non-centrosymmetric materials}}  (Gordon \& Breach, Philadelphia, 1992).
		
\bibitem{ivchenkopikus}
E.~L. Ivchenko, G.~E. Pikus.
{\textit{Superlattices and other heterostructures}} (Springer, 1997).

\bibitem{ivchenko05a}
E.~L. Ivchenko.
{\textit{Optical Spectroscopy of Semiconductor Nanostructures}} (Alpha
  Science, Harrow UK, 2005).
	
\bibitem{ganichev_book}
S.~Ganichev, W.~Prettl.
{\textit{Intense Terahertz Excitation of Semiconductors}} (Oxford University Press, 2006).

\bibitem{Ch7Yaroshetskii80p173} I.D.~Yaroshetskii and
S.M.~Ryvkin, \textit{The Photon Drag of Electrons in
Semiconductors} (in Russian), in Problems of Modern Physics  ed.
V.M. Tuchkevich and V.Ya. Frenkel (Nauka, Leningrad, 1980),
pp.~173-185 [English translation: {\it Semiconductor Physics}, ed.
V.M. Tuchkevich and V.Ya. Frenkel (Cons. Bureau, New York, 1986),
pp.~249-263].

\bibitem{Ch7Gibson80p182} A.F.~Gibson  and
M.F.~Kimmitt, \textit{ Photon Drag Detection}, in Infrared and
Millimeter Waves, Vol. 3, Detection of Radiation, ed. K.J.~Button
(Academic Press, New York, 1980), pp.181-217.

\bibitem{IvchenkoGanichev}
E.L. Ivchenko and S.D. Ganichev,
\textit{Spin Photogalvanics} in \textit{Spin Physics in Semiconductors}, ed. M.I. Dyakonov (Springer 2008) pp. 245-277. 

\bibitem{GanichevGolub} S.D. Ganichev and L.E. Golub,
phys. stat. solidi B - basic solid state physics \textbf{251}, 1801, (2014).

\bibitem{Saleh} B. E. A. Saleh, M. C. Teich, \emph{Fundamentals of Photonics} (John Wiley \& Sons, New York,  2003).

\bibitem{BornWolf} M. Born, E. Wolf, \emph{Principles of Optics: Electromagnetic Theory of Propagation, Interference and Diffraction of Light} (Cambridge University Press,  1999). 


\bibitem{DMS2009} S. D. Ganichev, S. A. Tarasenko, V. V. Bel'kov, P. Olbrich, W. Eder, D. R. Yakovlev, V. Kolkovsky, W. Zaleszczyk, G. Karczewski, T. Wojtowicz, and D. Weiss, 
Phys. Rev. Lett. \textbf{102}, 156602 (2009). 

\bibitem{Kvon2008}  Z.D. Kvon, S.N. Danilov, N.N. Mikhailov, S.A. Dvoretsky, and S.D.Ganichev, 
Physica E \textbf{40}, 1885 (2008).


\bibitem{Tunnel1993} S.~D.~Ganichev, W.~Prettl, and P.~G.~Huggard,
Phys. Rev. Lett.  {\bf 71}, 3882 (1993).

\bibitem{DX1995} S.~D.~Ganichev, I.~N.~Yassievich, W.~Prettl, J.~Diener,  B.~K.~Meyer and K.~W.~Benz,
Phys. Rev. Lett. {\bf  75},  1590 (1995).

\bibitem{spinrelax}  Petra~Schneider, J.~Kainz, S.D.~Ganichev, V.V.~Bel'kov,
S.N.~Danilov, M.M.~Glazov, L.E.~Golub, U.~R\"{o}ssler,
W.~Wegscheider, D.~Weiss, D.~Schuh, and W.~Prettl,
J. Appl. Phys.  {\bf 96}, 420 (2004).

\bibitem{Tunnelingfrequency1998} S.~D.~Ganichev, E.~Ziemann, Th.~Gleim, W.~Prettl, I.~N.~Yassievich, V.~I.~Perel, I.~Wilke, and E.~E.~Haller,
Phys. Rev. Lett. {\bf 80}, 2409 (1998).

\bibitem{FEL1} P. Michel, F. Gabriel, E. Grosse, P. Evtushenko, T. Dekorsy, M. Krenz, M. Helm,
U. Lehnert, W. Seidel, R.W\"{u}nsch, D.Wohlfarth, A.Wolf, 
Proceedings of the 2004 FEL Conference, 8-13.

\bibitem{FEL2} P. Michel, H. Buettig, F. Gabriel, M. Helm,
U. Lehnert, Ch. Schneider, R. Schurig, W. Seidel,
D. Stehr, J. Teichert, S. Winnerl, R. Wünsch, 
The Rossendorf IR-FEL ELBE, Proceedings of the 2006 FEL Conference, 488-491.


\bibitem{Knippels99p1578}  G.\,M.\,H. Knippels, X. Yan, A.\,M. MacLeod,
W.\,A. Gillespie, M. Yasumoto, D. Oepts, and A.\,F.\,G. van der Meer,
Phys. Rev. Lett. \textbf{ 83}, 1578 (1999). 

\bibitem{weber2008} W.~Weber, L.E.~Golub,  S.N. Danilov, J.~Karch, C.~Reitmaier, B.~Wittmann, V.V. Bel'kov,  
E.L.~Ivchenko, Z.D. Kvon, N.Q.~Vinh A.F.G.~van~der~Meer, B.~Murdin, and S.D.~Ganichev,
Phys. Rev. B \textbf{77}, 245304 (2008). 

\bibitem{Faist1994} J. Faist, F. Capasso, D. L. Sivco, C. Sirtori, A. L. Hutchinson, A. Y. Cho, 
Science \textbf{264}, 553 (1994). 

\bibitem{Faist2000} J. Faist, F. Capasso, C. Sirtory, D.L. Sivko, and A.Y. Cho,  \textit{Quantum Cascade Lasers}, in  series \textit{Semiconductors and Semimetals}, eds.R.K. Willardson and E.R. Weber, Vol. 66, \textit{Intersubband Transitions in Quantum Wells}, Volume eds. H.C. Liu and F. Capasso  (Academic Press, San Diego, 2000)

\bibitem{helm2001} M. Helm, \textit{Infrared long wavelength infrared emitters based on quantum wells and superlattices} (Gordon \& Breach Science Publishers, Amsterdam,  2000).

\bibitem{Bruenderman2012} E. Bruendermann, H.-W. Huebers, and M.F. Kimmitt,
\textit{Terahertz Techniques } in \textit{Springer Series in Optical Sciences},
(Springer-Verlag, Berlin and Heidelberg, 2012)


\bibitem{Ganichev1999} S. D. Ganichev
	Physica B \textbf{273-274}, 737 (1999).

	\bibitem{Ziemann2000}  E. Ziemann, S. D. Ganichev, I. N. Yassievich, V. I. Perel, and W. Prettl,
	J. Appl. Phys. {\bf 87}, 3843 (2000).
		

\bibitem{Bostwick09} A. Bostwick,
T. Ohta, T. Seyller, K. Horn, and E. Rotenberg,
Nature Phys. \textbf{3}, 36 (2007).

 
\bibitem{Virojanadara08} C. Virojanadara,
M. Syv\"ajarvi, R. Yakimova, L. I. Johansson, A. A. Zakharov, and T. Balasubramanian,
Phys. Rev. B \textbf{78}, 245403 (2008).

\bibitem{Emtsev2009}
K.~V. Emtsev, A.~Bostwick, K.~Horn, J.~Jobst, G.~L. Kellogg, L.~Ley, J.~L.
  McChesney, T.~Ohta, S.~A. Reshanov, J.~Rohrl, E.~Rotenberg, A.~K. Schmid,
  D.~Waldmann, H.~B. Weber, and T.~Seyller,
 {Nature Materials} {\bf 8}, 203 (2009).

\bibitem{Seyller2} 
M. Ostler, F. Speck, M. Gick, T. Seyller,
Phys. Stat. Sol. B \textbf{247}, 2924 (2010).

\bibitem{Tzalenchuk2010}
A.~Tzalenchuk, S.~Lara-Avila, A.~Kalaboukhov, S.~Paolillo, M.~Syvajarvi,
  R.~Yakimova, O.~Kazakova, J.~J.~B. M., V.~Fal'ko, and S.~Kubatkin,
{Nature Nanotechnology} {\bf 5}, 186 (2010).

\bibitem{suppllara2011} S.~Lara-Avila, K. Moth-Poulsen, R. Yakimova, T. Bjornholm, V. Fal'ko, A. Tzalenchuk, S. Kubatkin,
{Advanced Materials} \textbf{23}, 878 (2011).

\bibitem{SwedenPRL2011} S. Lara-Avila, A. Tzalenchuk, S. Kubatkin, R. Yakimova, T. J. B. M. Janssen, K. Cedergren, T. Bergsten, and V. Fal'ko,
Phys. Rev. Lett. \textbf{107}, 166602 (2011).




\bibitem{Bib:Novoselov2004} K.~S. Novoselov,
A. K. Geim, S. V. Morozov, D. Jiang,
Y. Zhang, S. V. Dubonos, I. V. Grigorieva, and A. A. Firsov,
Science {\bf 306}, 666 (2004).


\bibitem{perelpinskii73}
V.~I. Perel' and Ya.~M. Pinskii, 
\newblock {Sov. Phys. Solid State}, {\bf 15}, 688 (1973). 


\bibitem{Ivchenko1980}
E.~L. Ivchenko, G.~E. Pikus, in {\textit{Semiconductor Physics}} (Cons. Bureau, New York,  1986).

\bibitem{belinicher_cpde}
V.~I. Belinicher,
Sov. Phys. Solid State \textbf{23}, 2012 (1981).

\bibitem{Shalygin2007} 
V.~Shalygin, H.~Diehl, C.~Hoffmann, S.~Danilov, T.~Herrle, S.~Tarasenko,
  D.~Schuh, C.~Gerl, W.~Wegscheider, W.~Prettl, S.~Ganichev,
JETP Letters \textbf{84}, 570 (2007).

\bibitem{PhysRevLett.103.103906}
T.~Hatano, T.~Ishihara, S.~G. Tikhodeev, N.~A. Gippius,
 Phys. Rev. Lett. \textbf{103}, 103906 (2009).

\bibitem{3authors} S.D.~Ganichev, E.~L.~Ivchenko, and W.~Prettl,
Physica E {\bf 14}, 166 (2002).

\bibitem{Tarasenko2007}
S.~Tarasenko,
JETP Letters \textbf{85}, 182 (2007).

\bibitem{PhysRevB.79.121302}
P.~Olbrich, S.~A. Tarasenko, C.~Reitmaier, J.~Karch, D.~Plohmann, Z.~D. Kvon,
  S.~D. Ganichev, 
Phys. Rev. B \textbf{79}, 121302 (2009).

\bibitem{tarasenko11}
S.~A. Tarasenko,
 Phys. Rev. B \textbf{83}, 035313 (2011).

\bibitem{Borysiuk} 
J. Borysiuk, R. Bozek, W. Strupinski, A. Wysmolek, K. Grodecki, R. Stepniewski, and J. M. Baranowski,
J. Appl. Phys. \textbf{105}, 023503 (2009). 

\bibitem{Raman1} C. Casiraghi,
A. Hartschuh, H. Qian, S. Piscanec, C. Georgi, A. Fasoli, K. S. Novoselov, D. M. Basko, and A. C. Ferrari,
Nano Lett. \textbf{9}, 1433 (2009).

\bibitem{Raman3} S. Heydrich,
M. Hirmer, C. Preis, T. Korn, J. Eroms, D. Weiss, and C. Sch{\"u}ller, 
Appl. Phys. Lett. \textbf{97}, 043113 (2010).

\bibitem{SK_TS4} E.~J.~H Lee, K. Balasubramanian, R. Thomas Weitz, M. Burghard, 
and K. Kern,
Nature Nano. \textbf{3}, 486
(2008).

\bibitem{erl5} F. Speck, J. Jobst, F. Fromm, M. Ostler, D. Waldmann, M. Hundhausen, H. B. Weber, Th. Seyller,
Appl. Phys. Lett. \textbf{99}, 122106 (2011).

\bibitem{Falko89} V.~I. Falko, 
{Fiz. Tvedr. Tela} {\bf 31}, 29 (1989) [Sov. Phys. Solid State {\bf 31}, 561 (1989)].

\bibitem{Tarasenko08}
S.~A. Tarasenko, 
{Phys. Rev. B} {\bf 77}, 085328 (2008).

\bibitem{Tarasenko11}
S.~A. Tarasenko, 
{Phys. Rev. B} {\bf 83}, 035313 (2011).


\bibitem{ratchet2009} P. Olbrich, E. L. Ivchenko, T. Feil, R. Ravash, S. D. Danilov, J. Allerdings, D. Weiss, and S. D. Ganichev,
Phys. Rev. Lett. \textbf{103}, 090603 (2009). 
	
\bibitem{ratchet2011} P. Olbrich, J. Karch, E. L. Ivchenko, J. Kamann, B. M{\"a}rz, M. Fehrenbacher, D. Weiss, and S. D. Ganichev,
Phys. Rev. B \textbf{83}, 165320 (2011).

\bibitem{Review_JETP_Lett} E. L. Ivchenko and S. D. Ganichev,
JETP Lett. \textbf{93}, 673 (2011).

\bibitem{Theory_PRB_11}	A. V. Nalitov, L. E. Golub, and E. L. Ivchenko,
Phys. Rev. B \textbf{86}, 115301 (2012). 



\bibitem{footnoteLG} Here we assume the Boltzmann statistics for clarity. The case of the Fermi-Dirac statistics and real situation of degenerate carriers are studied in~\cite{ratchet_graphene16,Theory_PRB_11}.




\bibitem{footnoteT1}
Hereafter we consider a graphene sheet with the lateral potential ${\cal V}(x)$. 
The electron energy in each valley, $K$ or $K'$, is given by
%
$\varepsilon_{\bm k} = \hbar v_0 k + {\cal V}(x)$,
%
and the two-dimensional wave vector ${\bm k}$ is referred to the vortex of the hexagonal Brillouin zone. Since in the model under consideration the behavior of electrons in the $K$ or $K'$ valleys is identical we consider the current generation in one of them and then double the result. 

\bibitem{ellipticitydetector}
S.N. Danilov, B. Wittmann, P. Olbrich, W. Eder, W. Prettl, L.E. Golub, E.V.Beregulin, Z.D. Kvon, N.N. Mikhailov, S.A. Dvoretsky, V.A. Shalygin, N.Q. Vinh, A. F.G. van der Meer, B. Murdin, and S.D. Ganichev, J. Appl. Phys. \textbf{105}, 013106 (2009).

\bibitem{ellipticitydetector2} 
S. Dvoretsky, N. Mikhailov, Y. Sidorov, V. Shvets, S. Danilov, B. Wittman, and S. Ganichev, 
J. Electron. Mat. \textbf{39},  918 (2010). 





\bibitem{16w} V. V. Popov, 
J. Infr. Millim. THz Waves \textbf{32}, 1178 (2011).

\bibitem{31} V. V. Popov, D. V. Fateev, T. Otsuji, Y. M. Meziani, D. Coquillat, and W. Knap, 
Appl. Phys. Lett. \textbf{99}, 243504 (2011).  

\bibitem{115} E. S. Kannan, I. Bisotto, J.-C. Portal, T. J. Beck, and L. Jalabert, 
Appl. Phys. Lett. \textbf{101}, 143504 (2012).

\bibitem{otsuji2} S.A. Boubanga-Tombet, Y. Tanimoto, A. Satou, T. Suemitsu, Y. Wang, H. Minamide, H. Ito, D. V. Fateev, V.V. Popov, and T. Otsuji,
Appl. Phys. Lett. {\bf 104}, 262104 (2014).

\bibitem{Otsuji_Ganichev}  P. Faltermeier, P. Olbrich, W. Probst, L. Schell, T. Watanabe,
S. A. Boubanga-Tombet, T. Otsuji, and S. D. Ganichev,
J. Appl. Phys. \textbf{118}, 084301 (2015).


\bibitem{Ganichev84p20} S. D. Ganichev, Y. V. Terent'ev, and I. D. Yaroshetskii, 
	Pisma Zh. Tekh. Fiz. \textbf{11}, 46 (1985) [Sov. Tech. Phys. Lett. \textbf{11}, 20 (1989)].


\bibitem{tunnelreview02}S.D.~Ganichev, I.N.~Yassievich, and W.~Prettl,
J. Phys.: Condens. Matter {\bf 14}, R1263 (2002).

\bibitem{Cai}	X. Cai, A. B. Sushkov, R. J. Suess, M. M. Jadidi, G. S. Jenkins, L. O. Nyakiti, R. L. Myers-Ward, S. Li, J. Yan, D. K. Gaskill, T. E. Murphy, H. D. Drew and M. S. Fuhrer, 
Nat. Nanotechnol. \textbf{9}, 814 (2014).


\bibitem{HosurBerry} P. Hosur, 
Phys. Rev. B {\bf 83}, 035309 (2011).

\bibitem{TopIns4}J.W. McIver, D. Hsieh, H. Steinberg, P. Jarillo-Herrero, N. Gedik, 
Nat. Nanotechnol. {\bf 7}, 96 (2012).

\bibitem{Ultrathin} Quan Sheng Wu, Sheng Nan Zhang, Zhong Fang, Xi Dai, 
Physica E {\bf 44}, 895 (2012)

\bibitem{TopIns2}P. Olbrich, L.E. Golub, T. Herrmann, S.N. Danilov, H. Plank, V.V. Bel'kov, G. Mussler, Ch. Weyrich, C.M. Schneider, J. Kampmeier, D. Gr\"{u}tzmacher, L. Plucinski, M. Eschbach, S.D. Ganichev, 
Phys. Rev. Lett. {\bf 113}, 096601 (2014).

\bibitem{TopIns5} Junxi Duan, Ning Tang, Xin He, Yuan Yan, Shan Zhang, Xudong Qin, Xinqiang Wang, Xuelin Yang, Fujun Xu, Yonghai Chen, Weikun Ge, Bo Shen, 
Sci. Rep. {\bf 4}, 4889  (2014).

\bibitem{TItheory}V. Kaladzhyan, P.P. Aseev, S.N. Artemenko, 
Phys. Rev. B {\bf 92}, 155424 (2015).


\bibitem{TopIns3}K.-M. Dantscher, D.A. Kozlov, P. Olbrich, C. Zoth, P. Faltermeier, M. Lindner, G.V. Budkin, S. A. Tarasenko, V.V. Bel'kov, Z. D. Kvon, N.N. Mikhailov, S.A. Dvoretsky, D. Weiss, B. Jenichen, S.D. Ganichev, 
Phys. Rev. B {\bf 92}, 165314 (2015)

\bibitem{Ultrafast} Christoph Kastl, Christoph Karnetzky, Helmut Karl, A.W. Holleitner, 
Nat. Commun. {\bf 6}, 6617 (2015).

\bibitem{TopIns1}K.N. Okada, N. Ogawa, R. Yoshimi, A. Tsukazaki, K.S. Takahashi,
M. Kawasaki, Y. Tokura, 
Phys. Rev. B {\bf 93}, 081403 (2016).
%

\bibitem{TopIns2add} H. Plank, L.E. Golub, S. Bauer, V.V. Bel'kov, T. Herrmann, P. Olbrich, M. Eschbach, 
L. Plucinski, C.M. Schneider, J. Kampmeier, M. Lanius, G. Mussler, D. Gr\"{u}tzmacher, S.D. Ganichev, 
Phys. Rev. B {\bf 93}, 125434 (2016).

\end{thebibliography}
\end{document}